\documentclass[superscriptaddress,twocolumn,prb,showpacs,floatfix]{revtex4}
\usepackage[dvips]{graphicx}
\usepackage{amsmath,amsfonts,amssymb,bm,ulem}
\usepackage{graphicx}

\begin{document}

\title{Large Magnetoresistance of a Dilute $p$-Si/SiGe/Si Quantum Well in a Parallel Magnetic Field}

\author{I.L.~Drichko}
\affiliation{A.F. Ioffe Physico-Technical Institute of Russian
Academy of Sciences, 194021 St. Petersburg, Russia}
\author{I.Yu.~Smirnov}
\email{ivan.smirnov@mail.ioffe.ru}
\affiliation{A.F. Ioffe  Physico-Technical Institute of Russian
Academy of Sciences, 194021 St. Petersburg, Russia}
\author{A.V.~Suslov}
\affiliation{National High Magnetic Field Laboratory, Tallahassee,
FL 32310, USA}
\author{O.A.~Mironov}
\affiliation{Warwick SEMINANO R$\&$D Centre, University of Warwick
Science Park, Coventry CV4 7EZ, UK}
\author{D.R. Leadley}
\affiliation{Department of Physics, University of Warwick, Coventry,
CV4 7AL, UK}
\date{\today}
\begin{abstract}
We report the results of an experimental study of the
magnetoresistance  $\rho_{xx}$ in two samples of $p$-Si/SiGe/Si with
low carrier concentrations $p$=8.2$\times$10$^{10}$\,cm$^{-2}$ and
$p$=2$\times$10$^{11}$\,cm$^{-2}$. The research was performed in the
temperature range of 0.3-2 K in the magnetic fields of up to 18 T,
parallel to the two-dimensional (2D) channel plane at two
orientations of the in-plane magnetic field $B_{\parallel}$  against
the current $I$: $B_{\parallel} \perp I$ and $B_{\parallel}
\parallel I$. In the sample with the lowest density in the magnetic
field range of 0-7.2 T the temperature dependence of $\rho_{xx}$
demonstrates the metallic characteristics ($d \rho_{xx}/dT>$0).
However, at $B_{\parallel}$ =7.2 T the derivative $d \rho_{xx}/dT$
reverses the sign. Moreover, the resistance depends on the current
orientation with respect to the in-plane magnetic field. At
$B_{\parallel} \cong$ 13 T there is a transition from the dependence
$\ln(\Delta\rho_{xx} / \rho_{0})\propto B_{\parallel}^2$ to the
dependence $\ln(\Delta\rho_{xx} / \rho_{0})\propto B_{\parallel}$.
The observed effects can be explained by the influence of the
in-plane magnetic field on the orbital motion of the charge carriers
in the quasi-2D system.
\end{abstract}

\pacs{73.23.-b, 73.43.Qt} \maketitle

\section{Introduction} \label{intr}

Large positive magnetoresistance in two-dimensional systems in the
in-plane magnetic field is typically explained by a modification in
the spin system of the charge carriers. Studies of the in-plane
field induced magnetoresistance were conducted in numbers of
low-density heterostructures $n$-Si MOS~\cite{bib:1},
$n$-Si/SiGe~\cite{bib:2}, $n$-GaAs/AlGaAs~\cite{bib:3},
$p$-GaAs/AlGaAs~\cite{bib:3,bib:4,bib:5},
$n$-Al$_{0.4}$Ga$_{0.6}$As/AlAs/Al$_{0.4}$Ga$_{0.6}$As~\cite{bib:3}.
Indeed, within the narrow quantum well approximation the orbital
motion of the charge carriers is suppressed, but the Zeeman
splitting $g^* \mu_B B$ lifts spin degeneracy, where $g^*$ is the
effective g-factor, $\mu_B$ is the Bohr magneton, and $B$ is the
total magnetic field. Thus, when the magnetic field is rising the
conduction band is being split into two subbands with the opposite
spin directions and, finally, all spin-down electrons find
themselves in the bottom subband under the Fermi level.
Consequently, the electronic system becomes completely
spin-polarized. As a rule, in the magnetic fields exceeding a
critical value corresponding to the field of full spin polarization
a saturation of resistance is
observed~\cite{bib:1,bib:2,bib:3,bib:4}. Several spin effect-based
theories were developed to interpret the nature of positive
magnetoresistance~\cite{bib:6,bib:7,bib:8,bib:9}.

However, in some studies on $n$- and
$p$-GaAs/AlGaAs~\cite{bib:3,bib:5} no saturation of the
magnetoresistance was observed. Authors of the theory
Ref.~\onlinecite{bib:10} interpreted that experimental fact by
coupling of the in-plane field to the carrier orbital motion due to
the finite thickness of the 2D-layer. Validity of such a model
proves to be true experimentally: unlike cases of
Refs.~\onlinecite{bib:3,bib:5} where the investigated samples
$p$-GaAs/AlGaAs by all appearances had a \textit{wide} quantum well,
in $p$-GaAs/AlGaAs with the \textit{narrow} quantum well a distinct
saturation of positive magnetoresistance was observed~\cite{bib:4}.

It is worth noting that positive magnetoresistance in parallel
magnetic fields was observed in dilute carrier systems only.

In the system under study $p$-Si/Si$_{1-x}$Ge$_{x}$/Si a quantum
well formed in the strained Si$_{1-x}$Ge$_{x}$ is asymmetrical due
to only one of the silicon barriers to be boron doped. The threefold
(not considering a spin) degenerated  valence band of SiGe is split
into 3 subbands due to a strong spin-orbit interaction and a strain
[11]. The top subband occupied by heavy holes is formed by the
states with quantum numbers $L$=1, $S$=1/2, $J$=3/2. As a result,
there is a strong anisotropy of the g-factor with respect to the
magnetic field orientation: $g^*_{\perp} \cong$ 4.5 when a magnetic
field is applied perpendicular to the quantum well plane, and
$g^*_{\parallel} \cong$ 0 when the field is oriented in the plane of
the 2D channel~\cite{bib:12}. Thus, the Zeeman spin splitting in a
parallel field is small. Considering this fact several authors, for
example, ~\cite{bib:13,bib:14} stated that effect of an in-plane
magnetic field on the resistance in this material has to be very
weak. That opinion was based also on a few experiments~\cite{bib:14}
carried out on the p-Si/SiGe systems with
$p=$2.5$\times$10$^{11}$\,cm$^{-2}$. The experimental results
presented below show that a parallel magnetic field still greatly
affects the resistance in the low-density $p$-Si/SiGe/Si quantum
well despite the lack of the spin effect.

\section{Experiment and discussion} \label{experiment and discussion}

The DC magnetoresistance  $\rho_{xx}$ was measured in the in-plane
magnetic field on two $p$-Si/SiGe/Si quantum wells, with the carrier
densities $p$=8.2$\times$10$^{10}$\,cm$^{-2}$ and
$p$=2$\times$10$^{11}$\,cm$^{-2}$, respectively, and a hole mobility
of $\mu  = 1\times10^{4}$ cm$^2$/V$\cdot$s at liquid helium
temperatures. The studies were performed in the fields of up to 18 T
and in the temperature range of 0.3-2 K at two orientations of the
in-plane magnetic field $B_{\parallel}$ with respect to the current
$I$: $B_{\parallel} \perp I$ and $B_{\parallel} \parallel I$.

The 2D-system Si$\langle$B$\rangle$/Si/SiGe/ Si/(001)Si was grown in
Warwick on a Si (001) substrate by the solid source molecular beam
epitaxy. It consisted of the 300 nm Si buffer layer followed by 30
nm Si$_{0.92}$Ge$_{0.08}$ layer (sample with
$p$=8.2$\times$10$^{10}$\,cm$^{-2}$) or Si$_{0.87}$Ge$_{0.13}$ layer
(sample with $p$=2$\times$10$^{11}$\,cm$^{-2}$), 20 nm undoped
spacer and 50 nm layer of B-doped Si with doping concentration
2.5$\times 10^{18}$~cm$^{-3}$. Magnetoresistance of the sample with
$p$=8.2$\times$10$^{10}$\,cm$^{-2}$ was studied in detail earlier in
wide range of transverse magnetic fields and
temperatures~\cite{bib:15}. At the field $B$=0 T the resistance in
both samples demonstrated a metallic behavior.

\begin{figure}[t]
\centerline{
\includegraphics[width=8.8cm,clip=]{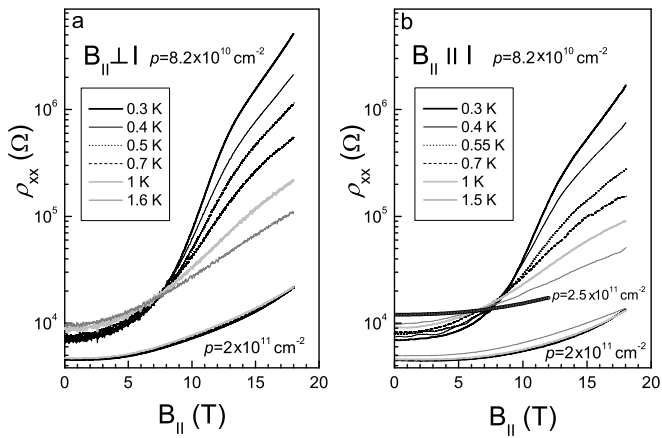}
} \caption{Magnetoresistance $\rho_{xx}$ versus $B_{\parallel}$
parallel to the two-dimensional system for the samples with
$p$=8.2$\times$10$^{10}$\,cm$^{-2}$ and
$p$=2$\times$10$^{11}$\,cm$^{-2}$ with two $B$-to-$I$
orientations: a) $B_{\parallel} \perp I$, b) $B_{\parallel}
\parallel I$; the curve for $p$=2.5$\times$10$^{11}$\,cm$^{-2}$ at
$T$=0.5 K is the result of Ref.~\onlinecite{bib:14}.\label{rxxB}}
\end{figure}

Fig.~\ref{rxxB} illustrates dependences of the resistance
$\rho_{xx}$ on the in-plane magnetic field $B_{\parallel}$ at
various temperatures.

It is evident, that $\rho_{xx} (B_{\parallel})$ curves related to
the sample with $p$=8.2$\times$10$^{10}$\,cm$^{-2}$ while been
measured at different temperatures cross at a single point
corresponding to $B_{\parallel} \approx$ 7.2 T and  $\rho_{xx} = 1.8
\times 10^4$ Ohm. At this field the resistivity does not depend on
the temperature, i.e. $d \rho_{xx} /d T$ changes from positive to
negative. Such crossing is often interpreted as a metal-to-insulator
transition. In other words, the metallic state is suppressed by
fields higher than 7.2 T. In the sample with
$p$=2$\times$10$^{11}$\,cm$^{-2}$ the increase of the
magnetoresistance is just three times in the field range of 0-18 T
(see Fig.~\ref{rxxB}). Over the entire range of the magnetic field
used in the experiments this sample shows a metallic behavior as $d
\rho_{xx} /d T >$0.

As seen in Fig.~\ref{rxxB} the resistance increase $[\rho_{xx}
(12\text{T})-\rho_{xx}(0)]/\rho_{xx}(0)\cong 50 \%$ was
observed~\cite{bib:14} on a sample with
$p$=2.5$\times$10$^{11}$\,cm$^{-2}$. That system also was in a
metallic state in the fields of up to $B_{\parallel}$=12 T used in
Ref.~\onlinecite{bib:14}.

\begin{figure}[h]
\centerline{
\includegraphics[width=6.9cm,clip=]{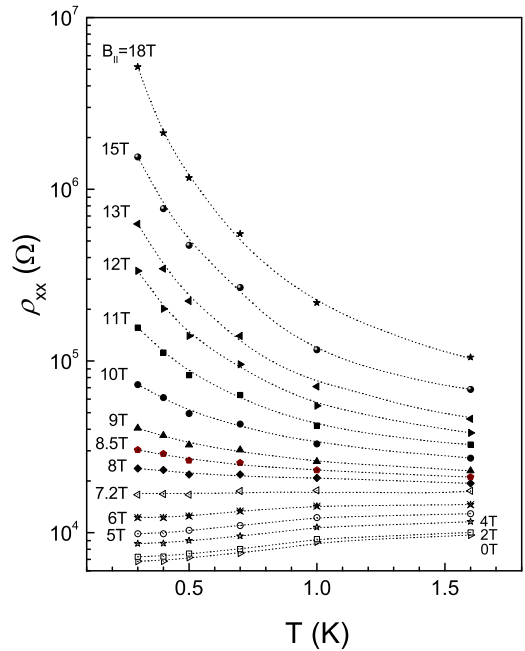}
} \caption{Resistivity $\rho_{xx}$ as a function of temperature $T$
for different magnetic fields for the sample with
$p$=8.2$\times$10$^{10}$\,cm$^{-2}$ in configuration $B_{\parallel}
\perp I$. \label{rxxBT}}
\end{figure}

In Fig.~\ref{rxxBT} the same magnetoresistance data obtained on the
sample with $p$=8.2$\times$10$^{10}$\,cm$^{-2}$ are replotted as the
dependencies  $\rho_{xx}(T)$ at several field values to clarify the
modification of the $d \rho_{xx} /d T$ and its sign change at the
field of 7.2 T.

The samples were mounted on a one axis rotator~\cite{bib:16}. That
allowed to determine the sample position at which the field was
aligned parallel to the quantum well plane with an error of less
than 10 minutes.

\begin{figure}[h]
\centerline{
\includegraphics[width=8.2 cm,clip=]{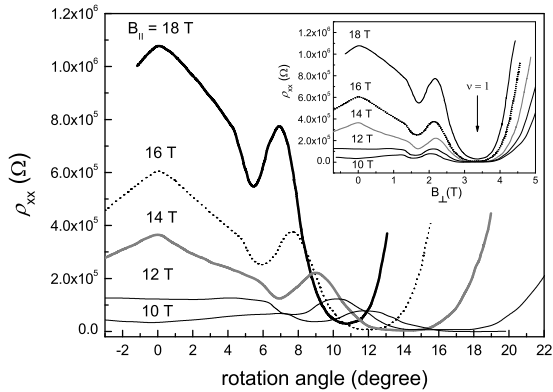}
} \caption{Resistance $\rho_{xx}$ as a function of the field tilting angle
$\Theta$ with respect to the plane of the 2D-layer at different
magnetic fields $B_{\parallel}$;
$p$=8.2$\times$10$^{10}$\,cm$^{-2}$, $B_{\parallel}
\parallel I$. The inset: $\rho_{xx}$ as a
function of a magnetic field $B_{\perp}$, the normal component of
the magnetic field.} \label{rotate}
\end{figure}

Fig.~\ref{rotate} illustrates the effect of sample rotation on the
resistivity $\rho_{xx}$. The angle $\Theta=0^{\text{o}}$
corresponding to the precise alignment of the magnetic field
parallel to the 2D-channel plane were determined at the $\rho_{xx}$
maximal value.

By rotating the sample in a constant $B_{\parallel}$, a
perpendicular field component is induced. That leads to appearance
of the Shubnikov -de Haas (SdH) oscillations. The oscillations
pattern allows tracking the concentration change with, for instance,
magnetic field intensity $B_{\parallel}$.  As seen in
Fig.~\ref{rotate}, with decrease of parallel component of applied
magnetic field the position of SdH oscillations shifts to larger
angles. However, if to replot $\rho_{xx}$ against the normal
component of magnetic field $B_{\perp}$ (see inset in
Fig.~\ref{rotate}) it appears that the oscillations minima for every
$B_{\parallel}$ are at the same $B_{\perp}$. If the positive
magnetoresistance is determined by spin effects, then in the tilted
field the structure of the SdH oscillations should depend on
intensity of the parallel field $B_{\parallel}$. If $B_{\parallel}$
is less than the field, corresponding to a totally polarized
electron system $B_c$, there is either appearance of oscillations
with another period~\cite{bib:3} or a change in $B_{\perp}$-position
of the oscillations minima, corresponding to the lower spin subband,
until the magnetic field achieves the point of total spin
polarization~\cite{bib:2}. These facts are indicative of the two
spin subbands repopulation with increase of the in-plane magnetic
field $B_{\parallel}$ till $B_c$.

According to our results (Fig.~\ref{rotate}) in the fields $B >$ 10
T nothing happens to position of the oscillations.  This may support
our interpretation that spin subbands are not formed.  Another
explanation would be that $B_c <$ 10 T.  The dependence $\rho_{xx}
(B_{\parallel})$ presented in Fig.~\ref{rxxB} shows pronounce
increase with magnetic field without any trend to saturation. Thus
explanation of such increase even assuming that $B > B_c$ would
require account of orbital effects.

\begin{figure}[h]
\centerline{
\includegraphics[width=8cm,clip=]{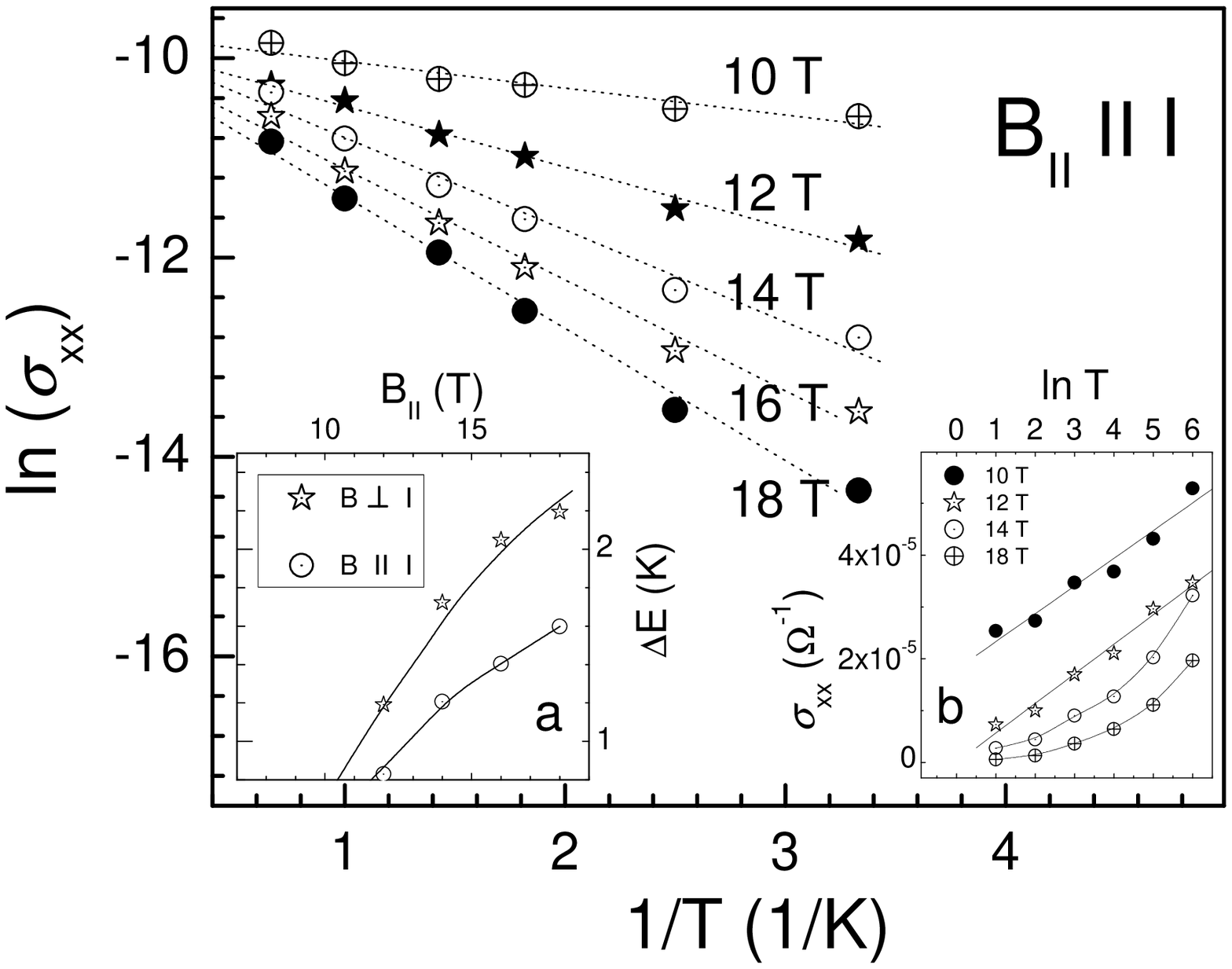}
} \caption{Arrhenius plot of the $p$-Si/SiGe conductivity at
different magnetic fields $B_{\parallel}$;
$p$=8.2$\times$10$^{10}$\,cm$^{-2}$. Insets: (a) Activation energy as a
function of in-plane magnetic field; (b) $\sigma_{xx}$ vs $\ln T$. } \label{lnsxx}
\end{figure}

The temperature dependence of the conductivity $\sigma_{xx}$ is
plotted in Fig.~\ref{lnsxx}. At $B>$12 T this dependence has an
activation character:  $\sigma_{xx} \propto \exp (-\Delta
E(B)/k_BT)$, with the activation energy  $\Delta E$ depending on the
in-plane magnetic field. The activation-like temperature dependence
at $B_{\parallel}>$ 12 T is a a characteristic of the hole
freezing-out on localized states. The corresponding conductance
mechanism could be the nearest neighbour hopping. The process of
magnetic freezing-out is associated with sufficiently high magnetic
fields where the wave function deformation takes place. The inset
(a) in Fig.\ref{lnsxx} shows that for different orientations this
deformation occurs in different ways: it is stronger for the
$B_{\parallel} \perp I$ configuration than for $B_{\parallel}
\parallel I$ one.

However, in the field region of 7-12 T the holes conductivity
mechanism could be hardly identified. Indeed, the temperature
dependence of conductivity could be interpreted as activation with
the energy $\Delta E \propto k_B T$ (see Fig.~\ref{lnsxx}), and
corresponding conductance mechanism could be the nearest neighbour
hopping. However, in the same magnetic fields region the condition
$\sigma_{xx}\propto \ln T$ is also valid (see inset (b) of
Fig.~\ref{lnsxx}), which could be expounded, for example, as quantum
corrections to conductivity of the delocalized holes as done in
Ref.~\onlinecite{bib:19}.

To specify the conductivity mechanisms of charge carriers (holes) in
parallel magnetic fields higher than 7.2 T it is useful to compare
the conductivity on a direct current with the high-frequency
conductivity. In the present work the AC conductivity, being
generally a complex function $\sigma^{\text{AC}} \equiv
\sigma_1-i\sigma_2$, has been measured on the same samples by
acoustic methods. The acoustic technique involves measurements of
the acoustoelectric effects: the variations of the velocity $\Delta
v/v(0)=(v (B)-v (0))/v (0)$ and the attenuation $\Delta
\Gamma=\Gamma (B) - \Gamma (0)$ of a surface acoustic wave (SAW)
($\Gamma (0)$, $v (0)$ are the absorption and velocity at $B$=0,
respectively) in the in-plane a magnetic field of up to 18 T at
temperatures 0.3-2 K. The measurements were performed in the
frequency range of $f=17-157$ MHz. Simultaneous measurements $\Delta
v/v(0)$ and $\Delta \Gamma$ allow to determine real $\sigma_1$ and
imaginary $\sigma_2$ components of the AC conductivity $\sigma^{AC}$
in the probeless way. As Ge and Si are not piezoelectric materials,
the hybrid method was used. According to this technique the SAW is
generated and propagates on a surface of a piezoelectric LiNbO$_3$
plate, while the sample is slightly pressed to the plate by springs.
Thus, the electric field, accompanying the SAW and having the same
frequency, penetrates the sample and interacts with the holes in a
quantum well. In this configuration of the experiment no deformation
is transmitted into the sample. AC measurements were also conducted
in two configurations: $k \parallel B_{\parallel}$ and $k \perp
B_{\parallel}$, where $k$ is the SAW wave vector.

\begin{figure}[h]
\centerline{
\includegraphics[width=6.6cm,clip=]{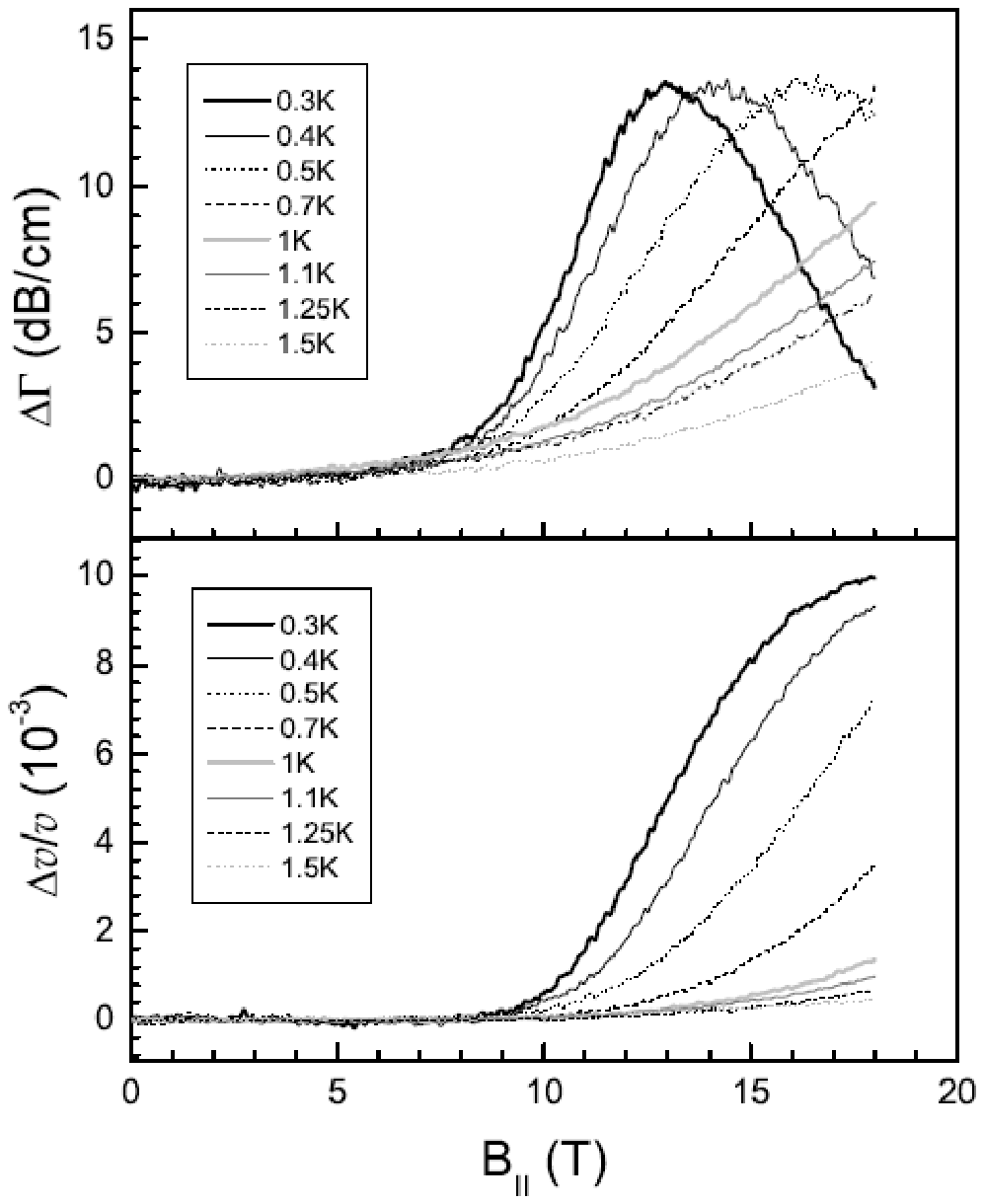}
} \caption{Magnetic field dependences of  $\Delta \Gamma (B_{\parallel})$ and $\Delta v(B_{\parallel})/v(0)$ for different temperatures.
$f$=17 MHz, $p=8.2 \times 10^{10}$cm$^{-2}$, $k \perp B_{\parallel}$.} \label{GV}
\end{figure}

Shown in Fig.~\ref{GV} are the parallel magnetic field dependences
of the acoustic attenuation, $\Delta \Gamma (B_{\parallel})$, and
velocity shift, $\Delta v(B_{\parallel})/v(0)$ at $f=17$ MHz at
different temperatures. The curves measured at other frequencies
(30, 90 and 157 MHz) are similar to the ones presented in
Fig.~\ref{GV}.

The components $\sigma_{1,2}$ can be found from simultaneous
measurements of $\Delta \Gamma$ and $\Delta v/v(0)$ by solving the
set of equations
\begin{eqnarray}
&&   \Delta \Gamma\,
  \propto \frac{\Sigma_1}{[1+\Sigma_2]^2+
  \Sigma_1^2},
\frac{\Delta v}{v(0)} \propto \frac{1+\Sigma_2}{[1+\Sigma_2]^2+
  \Sigma_1^2}, \nonumber \\
&&\Sigma_1 \propto \sigma_1,  \Sigma_2 \propto \sigma_2
\label{eq:GV}
\end{eqnarray}

The detailed procedure of determination of the components $\sigma_1$
and $\sigma_2$ from acoustic measurements is described in
Ref.~\onlinecite{bib:17}.

\begin{figure}[ht]
\centerline{
\includegraphics[width=\columnwidth]{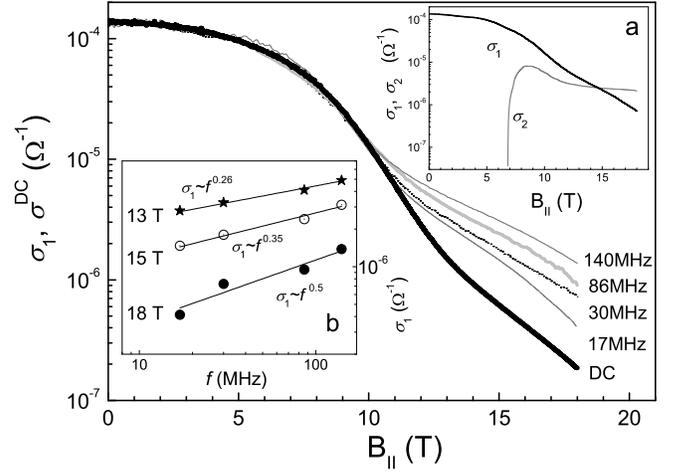}
}
\caption{Real part of the AC conductivity, $\sigma_1$ at different
  frequencies and DC conductivity versus in-plane magnetic
  field; $T=0.3$ K.  Inset (a): real, $\sigma_{1}$, and imaginary, $\sigma_{2}$, components of the
complex AC conductivity for $f$=86 MHz. Inset (b): frequency
  dependences of $\sigma_1$ at different magnetic fields, $T=0.3$ K.
 \label{fig:s1DCfreq}}
\end{figure}

Fig.~\ref{fig:s1DCfreq} illustrates experimental dependencies of the
real component $\sigma_{1}$ of the complex AC conductivity, derived
from the acoustical measurements at different frequencies using
Eqs.~(\ref{eq:GV}), as well as the DC-conductivity, on in-plane
field. Inset (a) displays the $\sigma_1$ and $\sigma_2$, measured at
frequency $f$=86 MHz as a function of in-plane field. Inset (b)
shows the frequency dependence of $\sigma_{1}$ at different magnetic
fields.

The ratio between $\sigma^{DC}$, $\sigma_1$ and $\sigma_2$ is
crucial in determination of the conductivity mechanism. If charge
carriers (holes) find themselves in delocalized states the AC and DC
conductivity mechanisms are the same, and $\sigma^{DC} = \sigma^{AC}
= \sigma_1$, while the imaginary part $\sigma_2=$0~\cite{bib:18}. It
is seen in Fig.~\ref{fig:s1DCfreq}, that AC and DC conductivities
coincide in the fields of up to $B_{\parallel} \approx$ 11-12 T. As
is also obvious from the figure, the magnetic field, at which
$\sigma^{DC}$ and $\sigma_1$ start to diverge, depends on a
frequency: the higher the frequency the less this field.

The AC and DC conductivity up to $B_{\parallel} \approx$ 11-12 T is
due to delocalized holes. At $B_{\parallel}>$ 12 T a localization of
the charge carriers begins, resulting in the nearest neighbour
hopping in the DC and the "two-site" hopping AC
conductivity~\cite{bib:20}. The latter is characterized by a
frequency dependence of $\sigma_1$ (see inset (b) in
Fig.~\ref{fig:s1DCfreq}) and emergence of the imaginary component
$\sigma_2$ (see inset (a) in Fig.~\ref{fig:s1DCfreq}). Moreover, as
seen in inset (a) of Fig.~\ref{fig:s1DCfreq}, at fields
$B_{\parallel}>$ 15 T the ratio $\sigma_2 > \sigma_1$ is valid,
which is typical of the "two-site" hopping for two-dimensional
systems~\cite{bib:21}.

According to these results, at $B_{\parallel} >$ 12 T  the charge
carriers are localized and the conductance is due to hopping.  It
supports the conclusion that near $B_{\parallel} =$7.2 T the hole
states are seemingly extended up to $B_{\parallel} \approx$12 T.
Consequently, temperature dependence of conductivity is most likely
determined by interplay between weak localization and  e-e
interactions. In this paper we do not focus on detailed study of the
conductivity in the vicinity of $B_{\parallel}=$7.2 T because of the
small variation of conductivity with temperature  and narrow
available temperature interval.
\begin{figure}[h]
\centerline{
\includegraphics[width=6.6cm,clip=]{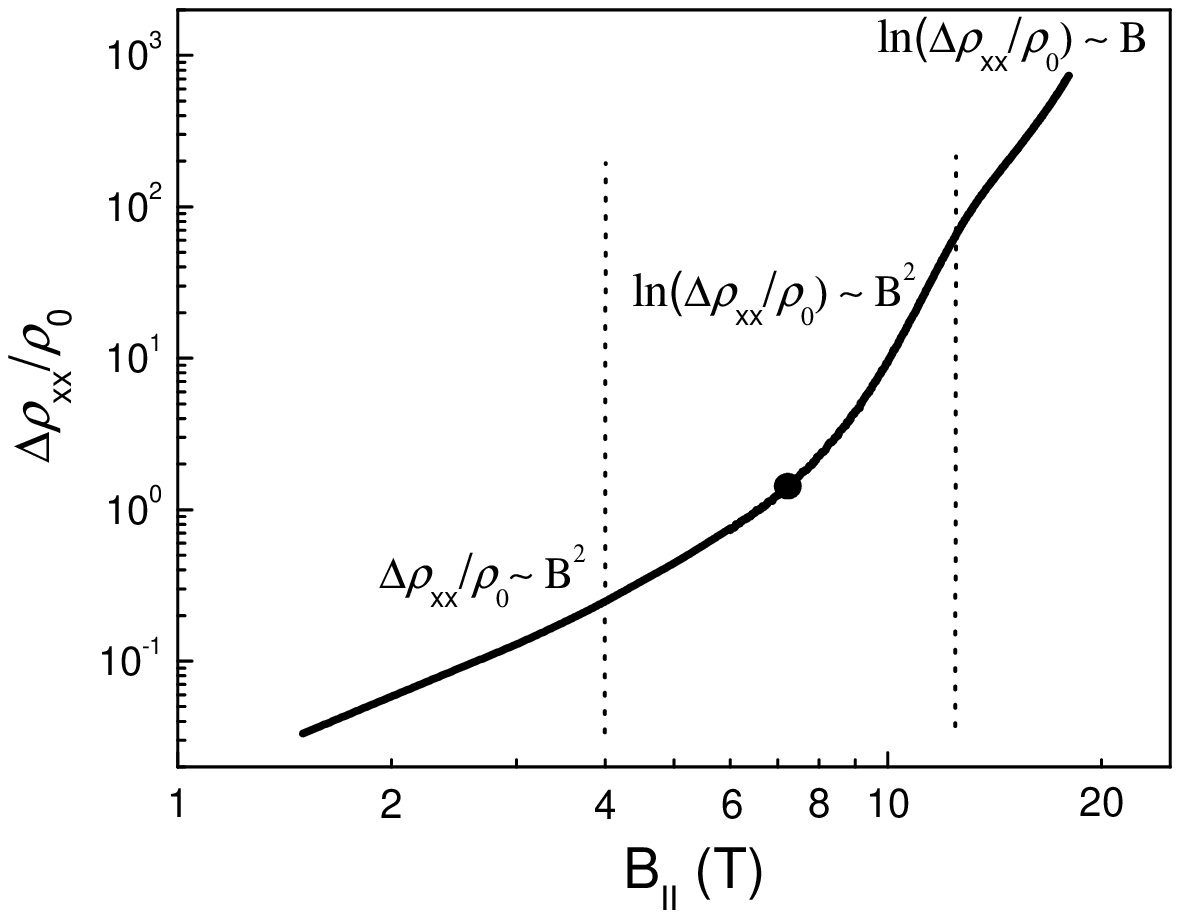}
} \caption{Dependence of $\Delta \rho_{xx} / \rho_{0}$ on
$B_{\parallel}$ at $T$=0.3 K for the sample with
$p$=8.2$\times$10$^{10}$\,cm$^{-2}$. $B_{\parallel} \perp I$.}
\label{rxxregions}
\end{figure}

We now consider the magnetic field dependence of the
magnetoresistance shown in Fig.\ref{rxxregions}. The single point on
the curve corresponds to $B_{\parallel}$=7.2 T. On the metallic side
until $B_{\parallel}$=4 T the magnetoresistance obeys a power law
$\Delta\rho_{xx} / \rho_{0} \propto B^2$. In the fields below 13 T
the resistivity follows a law $\ln (\Delta \rho_{xx} / \rho_{0}
)\propto B^2$, but then at $B_t \cong$ 13 T a transition to a
dependence $\ln (\Delta \rho_{xx} / \rho_{0})\propto B$ occurs.

\section{Summary} \label{Summary}

The principal experimental facts revealed in this study are the
following:

1)  An in-plane magnetic field induces a giant positive
magnetoresistance (with no sign of saturation) in $p$-Si/SiGe
heterostructures. This effect is observed in the system with lowest
density ($p$=8.2$\times$10$^{10}$\,cm$^{-2}$) only. When the
magnetic field rises up to 18 T, the resistivity in this sample
increases significantly and changes by more than three orders of
magnitude, while in the sample with
$p$=2$\times$10$^{11}$\,cm$^{-2}$ the magnetoresistance increase is
just several times in the same field range.

2) In the sample with $p$=8.2$\times$10$^{10}$\,cm$^{-2}$ the
derivative $d \rho_{xx}/d T$ changes its sign at $B \cong$ 7.2 T. It
could be interpreted as a magnetic field suppression of the metallic
state. In the sample with $p$=2$\times$10$^{11}$\,cm$^{-2}$ the sign
change of $d \rho_{xx}/d T$ was not observed.

3) The magnetoresistance in $p$-Si/SiGe heterostructure with
$p$=8.2$\times$10$^{10}$\,cm$^{-2}$ is anisotropic: it depends on
orientation of the current with respect to the in-plane magnetic
field, $B_{\parallel} \perp I$ or $B_{\parallel} \parallel I$. At
$B_{\parallel}=$ 18 T and $T$=0.3 K the resistivity ratio
$\rho_{xx}^{B \perp I} / \rho_{xx}^{B
\parallel I}$ equals 3.

4) In the sample with $p$=8.2$\times$10$^{10}$\,cm$^{-2}$ at $B_t
\cong$ 13 T a transition from the law $\ln (\Delta \rho_{xx} /
\rho_{0} )\propto B^2$ to $\ln (\Delta \rho_{xx} / \rho_{0})\propto
B$ was observed.

All these experimental facts could be explained qualitatively using
the theory developed in Ref.~\onlinecite{bib:10}. Authors of
Ref.~\onlinecite{bib:10} neglect spin-related effects and consider
the in-plane magnetoresistance as affected by the charge carrier
orbital motion in the wide quantum well.

The theory~\cite{bib:10} is applicable if the following two
conditions are met:

(a) the magnetic length $l_B=\sqrt{\hbar c/ eB}$ is less than the 2D
layer thickness $Z$: $Z \gg l_B$.

If to assume that the quantum well bottom is distorted weakly, one
may hold $Z \sim 3 \times 10^{-6}$ cm. Then the magnetic length
$l_B$ becomes less than thickness of 2D layer and condition (a) is
fulfilled already at $B_{\parallel} \cong$ 1 T.

(b) the 2D Fermi wavelength  $l_F$ at $B$=0 is substantially larger
than the magnetic length  $\lambda_F \gg l_B$, where
$\lambda_F=\frac{2 \pi}{k_F}$; $k_F= (2 \pi p)^{1/2}$, which is also
valid in our experiments.

The theory~\cite{bib:10} predicts (i) a large positive
magnetoresistance in systems with low carrier density; (ii) a large
anisotropy of resistivity in the 2D plane  $\rho^{\perp} \gg
\rho^{\parallel}$; (iii) a reduction of the effect with increasing
density; and (iv) a change of the $\rho (B)$ dependence: at low B
($\omega_c < \omega_0$), $\ln(\rho_{xx}) \propto B^2$, at high B
($\omega_c > \omega_0$), $\ln(\rho_{xx}) \propto B$, where
$\hbar\omega_0$ is the subband splitting,  and  $\omega_c$  is the
cyclotron frequency. All these predictions (i) - (iv) seem to be
consistent with the obtained experimental data.

Thus, one can calculate the thickness of the 2D well if the $B_t
\cong$ 13 T value is known. The transition occurs when
\begin{equation}
  \label{eq:01}
  \omega_0 = \omega_c:  \omega_0=\frac{\Delta E}{\hbar} = \frac{(E_2-E_1)}{\hbar}; \omega_c=\frac{e B}{m^*
  c},
\end{equation}
where $E_1$, $E_2$ are the energies of 1st and 2nd level of
dimensional quantization, $E_n=\frac{\hbar^2 \pi^2}{2m^*}
\frac{n^2}{Z^2}$, $n$ is the number of the dimensional quantization
band

It follows from Eq.~(\ref{eq:01}) that:
\begin{equation}
  \label{eq:02}
  Z^2=(n^2_2-n^2_1)(\pi^2 \hbar c)/(2eB_t).
\end{equation}

A calculation with Eq.~(\ref{eq:02}) gives $Z$=2.7$\times$10$^{-6}$
cm, which is consistent with the nominal thickness of the SiGe layer
in the studied 2D hole system: $Z$=3$\times$10$^{-6}$ cm.

It should be noted that Eq.~(\ref{eq:02}) provides only a rough
estimation of $Z$ because a field dependence of $E_n$ is not taken
into account there.

In conclusion, the large positive magnetoresistance in the parallel
magnetic field increases most likely due to interaction of the
in-plane field with the carrier orbital motion in the quasi-2D layer
with finite thickness and can be qualitatively explained by the
theory~\cite{bib:10}. The absence of a strong effect of the in-plane
magnetic field on the resistivity in Ref.~\onlinecite{bib:14} can be
possibly attributed to the relatively high carrier density in the
system studied there. This is consistent with our data obtained on
the sample with $p$=2$\times$10$^{11}$\,cm$^{-2}$.

\acknowledgments

Authors are grateful to Y.M. Galperin, L. Golub, M. Glazov, G.
Min'kov for useful discussions and to E. Palm and T. Murphy for
their help with the experiments. This work was supported  by grants
of RFBR 08-02-00852; the Presidium of the Russian Academy of
Science, the Program of Branch of Physical Sciences of RAS
"Spintronika", by the NSF through Cooperative Agreement No.
DMR-0084173, the State of Florida, and the DOE.

\vfill\eject


\begin{thebibliography}{11}

\bibitem{bib:1}
D. Simonian, S.V. Kravchenko, M.P. Sarachik, V.M. Pudalov, \prl
\textbf{79}, 2304 (1997), A.A. Shashkin, E.V. Deviatov, V.T.
Dolgopolov, A.A. Kapustin, S. Anissimova, A. Venkatesan, S.V.
Kravchenko, T.M. Klapwijk, \prb \textbf{73}, 115420 (2006).


\bibitem{bib:2}
T. Okamoto, M. Ooya, K. Hosoya, S. Kawaji, \prb \textbf{69}, 041202
(2004); K. Lai, W. Pan, D.C. Tsui, S.A. Lyon, M. Muhlberger, and F.
Schaffler, \prb \textbf{72}, 081313(R) (2005).


\bibitem{bib:3}
E. Tutuc, S. Melinte, and M. Shayegan, \prl \textbf{88}, 036805
(2002); E. Tutuc, E.P. De Poortere, S.J. Papadakis and M. Shayegan,
\prl \textbf{86}, 2858 (2001); E. Tutuc, E.P. De Poortere, S.J.
Papadakis and M. Shayegan, Physica E, 13,748 (2002).


\bibitem{bib:4} X.P.A. Gao, G.S. Boebinger, A.P. Mills Jr.,  A.P. Ramirez, L.N.
Pfeiffer, and K.W. West, \prb \textbf{73}, 241315(R) (2006).


\bibitem{bib:5}
J. Yoon, C.C. Li, D.Shahar, D.C. Tsui and M. Shayegan, \prl
\textbf{84}, 4421 (2000).



\bibitem{bib:6}
V.T. Dolgopolov, A. Gold, JETP Lett. \textbf{71}, 27 (2000) [Pis'ma
v ZhETF \textbf{71}, 42 (2000)].



\bibitem{bib:7}
B. Spivak, \prb \textbf{67}, 125205 (2003).


\bibitem{bib:8}
P. Phillips, Yi Wan, I. Martin, S. Knysh, D. Davidovich, Nature
{395}, 253 (1998).

\bibitem{bib:9}
G. Zala, B.N. Narozhny, I.L. Aleiner, \prb \textbf{64}, 214204
(2001).



\bibitem{bib:10}
S. Das Sarma and E.H. Hwang, \prl \textbf{84}, 5596(2000).


\bibitem{bib:11}
T.P. Pearsall, F.H. Pollak, J.C. Bean and R. Hull, \prb \textbf{33},
6821 (1986).

\bibitem{bib:12}
E. Glaser, J.M. Trombetta, T.A. Kennedy, S.M. Prokes, O.J.
Glembocki, K.L. Wang and C.H. Chern, \prl \textbf{65}, 1247
(1990).

\bibitem{bib:13} E. Abrahams,  S.V. Kravchenko, M.P. Sarachik, \rmp \textbf{73}, 251
(2001).

\bibitem{bib:14} S.I. Dorozhkin, C.J. Emeleus, T.E. Whall, G. Landwehr, \prb
\textbf{52}, R11638 (1995).

\bibitem{bib:15}
I.L. Drichko, A.M. Dyakonov, I.Yu. Smirnov, A.V. Suslov, Y.M.
Galperin, V. Vinokur, M. Myronov,  O.A. Mironov, D.R. Leadley, \prb
\textbf{77}, 085327 (2008).

\bibitem{bib:16}
E.C. Palm and T.P Murphy, Rev. Sci. Instr. \textbf{70}, 237
(1999).

\bibitem{bib:19}
X.P.A. Gao, A.P. Mills, Jr., A.P. Ramirez, L.N. Pfeiffer, K.W. West,
Phys. Rev. Lett. \textbf{89}, 016801 (2002); Phys. Rev. Lett.
\textbf{88}, 166803 (2002).

\bibitem{bib:17}
I.L. Drichko, A.M. Diakonov, I.Yu. Smirnov, Yu.M. Galperin, A.I. Toropov, \prb \textbf{62}, 7470 (2000).




\bibitem{bib:18}
I.L. Drichko, A.M. Diakonov, A.M. Kreshchuk, T.A. Polyanskaya, I.G.
Savel'ev, I.Yu. Smirnov, A.V. Suslov, FTP \textbf{31}, 451 (1997) [
Semiconductors \textbf{31}, 384 (1997)].



\bibitem{bib:20}
M. Pollak, T.H. Geballe, Phys. Rev. \textbf{122}, 1742 (1961).

\bibitem{bib:21}
A.L. Efros, Zh. Eksp. Teor. Phys. \textbf{89}, 1834 (1985) [JETP
\textbf{89}, 1057 (1985)].


\end{thebibliography}
\end{document}